\documentclass[twoside]{dis07}
\usepackage[latin1]{inputenc}
\usepackage[dvips]{graphicx,epsfig,color}
\usepackage{wrapfig,rotating}
\usepackage{amssymb,amsmath,array}

\begin{document}
\title{Transversity Signals in Two-Hadron Production at COMPASS}

\author{Christian Schill on behalf of the COMPASS collaboration
%
%
\vspace{.3cm}\\
%
Physikalisches Institut der Albert-Ludwigs-Universit\"at Freiburg \\
Hermann-Herder-Str. 3, 79104 Freiburg, Germany.
}

\maketitle

\begin{abstract}
New results on single spin asymmetries of identified charged pion and kaon pairs
produced in deep-inelastic scattering of muons on a transversely polarized
$^6LiD$ target are presented. The data were taken in the years 2003 and 2004
with the COMPASS spectrometer at CERN with a $160$~GeV muon beam from the CERN SPS
accelerator. The asymmetries can be interpreted in the context of transversity
as a convolution of the chiral-odd interference fragmentation function 
$H_1^\sphericalangle$ with the transverse spin distribution of quarks 
$\Delta_Tq(x)$. The measured azimuthal target spin asymmetries on 
the deuteron are compatible with zero within a small statistical error 
of about 1\%.
\end{abstract}

\section{Introduction}

An important missing piece in our understanding of the spin structure of the
nucleon is the transversity distribution function $\Delta_Tq(x)$. It is the only
one of the three leading-twist quark distribution functions $q(x), \Delta q(x)$
and $\Delta_Tq(x)$ that so-far remains unmeasured. The function  $\Delta_Tq(x)$
describes the distribution of transversely polarized  quarks in a transversely
polarized nucleon. It is difficult to measure $\Delta_Tq(x)$, since it is a
chiral-odd function which can only be probed in combination with another
chiral-odd function. So far, attempts were made to access transversity in
convolution with the Collins fragmentation function in single hadron production
\cite{Bressan, COMPASS}.

An alternative probe suggested to access transversity is the measurement of
two-hadron production in semi-inclusive deep-inelastic scattering on a
transversely polarized target. In this case, transversity is accessible via the
chiral-odd two-hadron interference fragmentation function
$H_1^\sphericalangle(z,M_h^2)$. The properties of interference fragmentation
functions are described in detail in Refs. \cite{Collins, Artru, Jaffe,
Bianconi, Radici, Bacchetta}. New COMPASS results for identified pion and kaon pairs are
presented in this contribution \cite{url}.

\section{Two-hadron asymmetry}

\begin{wrapfigure}{r}{0.5\columnwidth}
\centerline{\includegraphics[width=0.45\columnwidth]{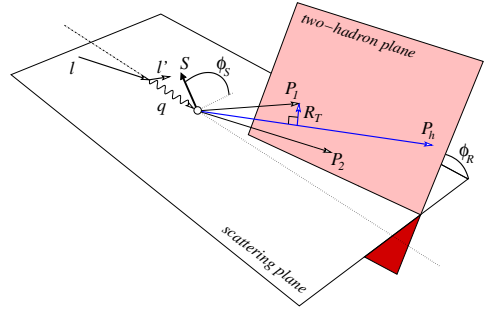}}
\caption{Definition of azimuthal angles $\phi_R$ and $\phi_S$.}\label{angles}
\end{wrapfigure}

At leading twist, the fragmentation function of a polarized quark into a pair of
hadrons is expected to be of the form \\[-0.8cm]

\begin{equation*}
D_q^{2h}(z, M_h^2) +  H_1^\sphericalangle(z,M_h^2) \sin\theta \sin \phi_{RS},
\end{equation*}
where $M_h$ is the invariant mass of the hadron pair and $z=z_1+z_2$ is the
fraction of available energy carried by the two hadrons. $D_q^{2h}(z, M_h^2)$ is
the unpolarized fragmentation function into two hadrons, and the interference
fragmentation function $H_1^\sphericalangle(z,M_h^2)$
is the spin dependent T-odd part of the fragmentation function of a transversely polarized quark
$q$ into a hadron pair. The angles $\theta$ and $\phi_{RS}$ (see Fig.
\ref{angles}) are defined according to Ref. \cite{Artru2}. 
$\phi_{RS}=\phi_R+\phi_{S}-\pi$ is the sum  of the azimuthal angle $\phi_R$ of
a plane containing the two hadrons and the  azimuthal angle $\phi_S$ of target
spin vector with respect to the lepton scattering plane. In COMPASS $\phi_R$ is defined as the azimuthal angle of
$\overrightarrow{R}_T$, the
transverse component of the vector
\begin{equation*}
\overrightarrow{R}=\frac{z_2\cdot \overrightarrow{p}_1-z_1\overrightarrow{p}_2}
{z_1+z_2}
\end{equation*}
where the indices 1 and 2 refer to the two final state hadrons. $\theta$ is the
polar angle of the first hadron in the two-hadron center-of-mass frame with
respect to the direction of the summed hadron momentum
$\overrightarrow{p}_h=\overrightarrow{p}_1+\overrightarrow{p}_2$. With the
applied cuts, $\theta$ peaks close to $\pi/2$ with $<\sin\theta> \approx 0.95$.
The following results are obtained by integrating over $\sin\theta$. 
The number of hadron pairs in a bin of the Bjorken variable $x$, or of
$z$, or of $M_h$ is given by 
\begin{equation*}
N^\pm(\phi_{RS})=N_0\cdot(1\pm A_{UT}^{\sin\phi_{RS}}\cdot \sin \phi_{RS}),
\end{equation*}
where $\pm$ refers to the transverse target spin orientation and $N_0$ is the
mean number of detected hadron pairs averaged over $\sin\phi_{RS}$. 
From the angular distribution of the hadron pairs, one can thus measure the
asymmetry 
\begin{equation*}
A_{RS}=\frac{1}{f P_T D}\cdot A_{UT}^{\sin\phi_{RS}},
\end{equation*}
where $f\approx 0.38$ is the target dilution factor, $P_T\approx 0.5$ 
the target polarization and $D$ the depolarization factor given by
 $D=(1-y)/(1-y+y^2/2)$, where  $y$ is the fractional energy
transfer of the lepton.

The measured asymmetry can be factorized into a convolution of the transversity
distribution $\Delta_Tq(x)$ of the quarks of flavor $q$ and the interference
fragmentation  $H_1^\sphericalangle(z,M_h^2)$ \cite{Artru2}:
\begin{equation*}
A_{RS}=\frac{\Sigma_q \;e_q^2 \cdot \Delta_Tq(x) \cdot H_1^\sphericalangle(z,M_h^2)}
{\Sigma_q \;e_q^2 \cdot q (x) \cdot D_q^{2h}(z, M_h^2)},
\end{equation*}
summed over all quark flavors $q$. Both the interference fragmentation function
$H_1^\sphericalangle(z,M_h^2)$ and the corresponding spin averaged fragmentation
function into two hadrons $D_q^{2h}(z, M_h^2)$ are unknown, and need to be measured in $e^+e^-$
annihilation or to be evaluated using models \cite{Jaffe, Bianconi, Radici, Bacchetta}. 
They are expected to depend on
$z=z_1+z_2$ and on the invariant mass $M_h$ of the two hadrons. 

\section{Event selection} The data discussed here have been taken in the years 2003
and 2004 in the COMPASS experiment  at CERN \cite{COMPASS}. It uses a secondary
$160$ GeV $\mu^+$ beam from $\pi$ and $K$ decay in the CERN SPS M2 beamline. The muons
are scattered on a  transversely polarized solid state $^6LiD$ target. The
target consists of two $60$~cm long target cells with opposite polarization. To
minimize systematic effects, the direction of the target polarization is reversed in
both cells once a week. The scattered muons and the produced hadrons
are detected in a $50$~m long large-acceptance forward spectrometer with
excellent particle identification capabilities. A large scale Ring Imaging
Cherenkov detector (RICH) is used to distinguish pions, kaons and protons
\cite{Experiment}. It allows to separate pions and kaons over a large momentum
range from the Cherenkov thresholds of $3$~GeV/c respectively $9$~GeV/c for pions and
kaons up to $43$~GeV/c.

The event selection was done in the same way as in the previous analysis of the
Collins and Sivers asymmetries \cite{Bressan, COMPASS}. For the selection of
the DIS event sample, kinematic cuts of the squared four-momentum transfer
$Q^2> 1$~(GeV/c)$^2$, the hadronic invariant mass $W> 5$~GeV/c$^2$ and the
fractional energy transfer of the muon $0.1 <y <0.9$ were applied. The average
$x$, $y$ and $Q^2$ in the final data sample is $x=0.035$, $y=0.33$ and
$Q^2=2.4$~(GeV/c)$^2$. The mean hadron multiplicity is $1.9$ hadrons per event.

Hadron pairs originating from the primary vertex are selected, where the first
hadron has positive and the second one negative charge. The hadron pairs are
separated into $\pi^+\pi^-$, $K^+\pi^-$, $\pi^+K^-$, and $K^+K^-$ pairs based
on the information of the RICH detector. Selection cuts of $z_1>0.1$ and
$z_2>0.1$ suppress hadrons from the target fragmentation and $z=z_1+z_2<0.9$ 
reject exclusively produced $\rho$-mesons. The resulting event sample contains
$3.7$~million $\pi^+\pi^-$, $0.3$~million $K^+\pi^-$, $0.25$~million $\pi^+K^-$
and $0.1$~million $K^+K^-$ pairs. By combining data from both target cells as well as
from sub-periods with opposite target polarization in a double ratio product
described in detail in Ref. \cite{COMPASS}, the acceptance function of the
spectrometer cancels out and the azimuthal asymmetry $A_{RS}(x, z, M_h) $ is 
extracted by a fit to the data. In various studies, it could be shown that 
systematic effects of the measurement are considerably smaller than the
statistical uncertainty of the data.

\section{Results} 
\begin{figure}[t]
\vspace{-0.1cm}\centerline{
\includegraphics[width=0.36\textwidth,height=3.1cm]{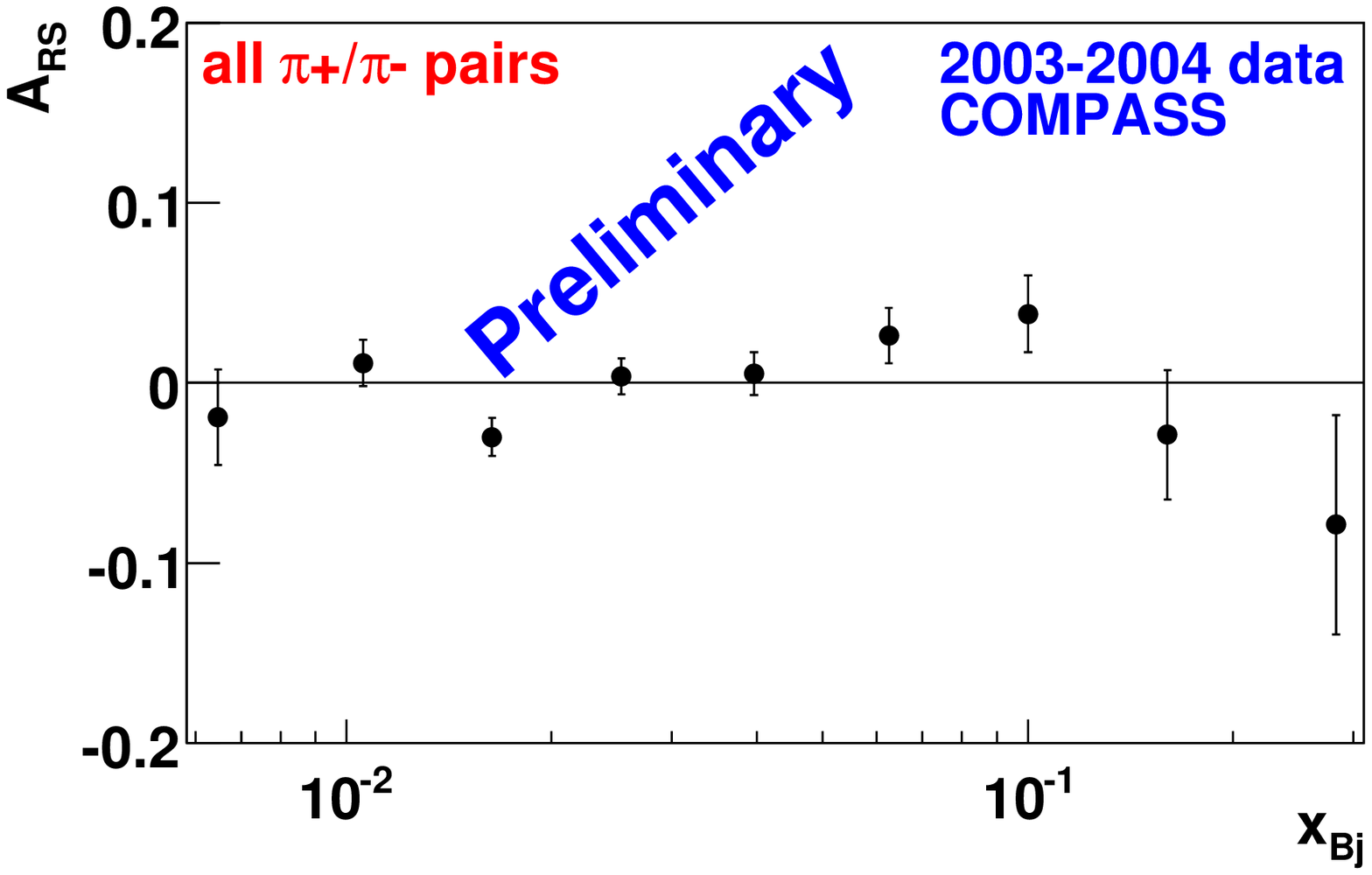}\hspace{-0.5cm}
\includegraphics[width=0.36\textwidth,height=3.1cm]{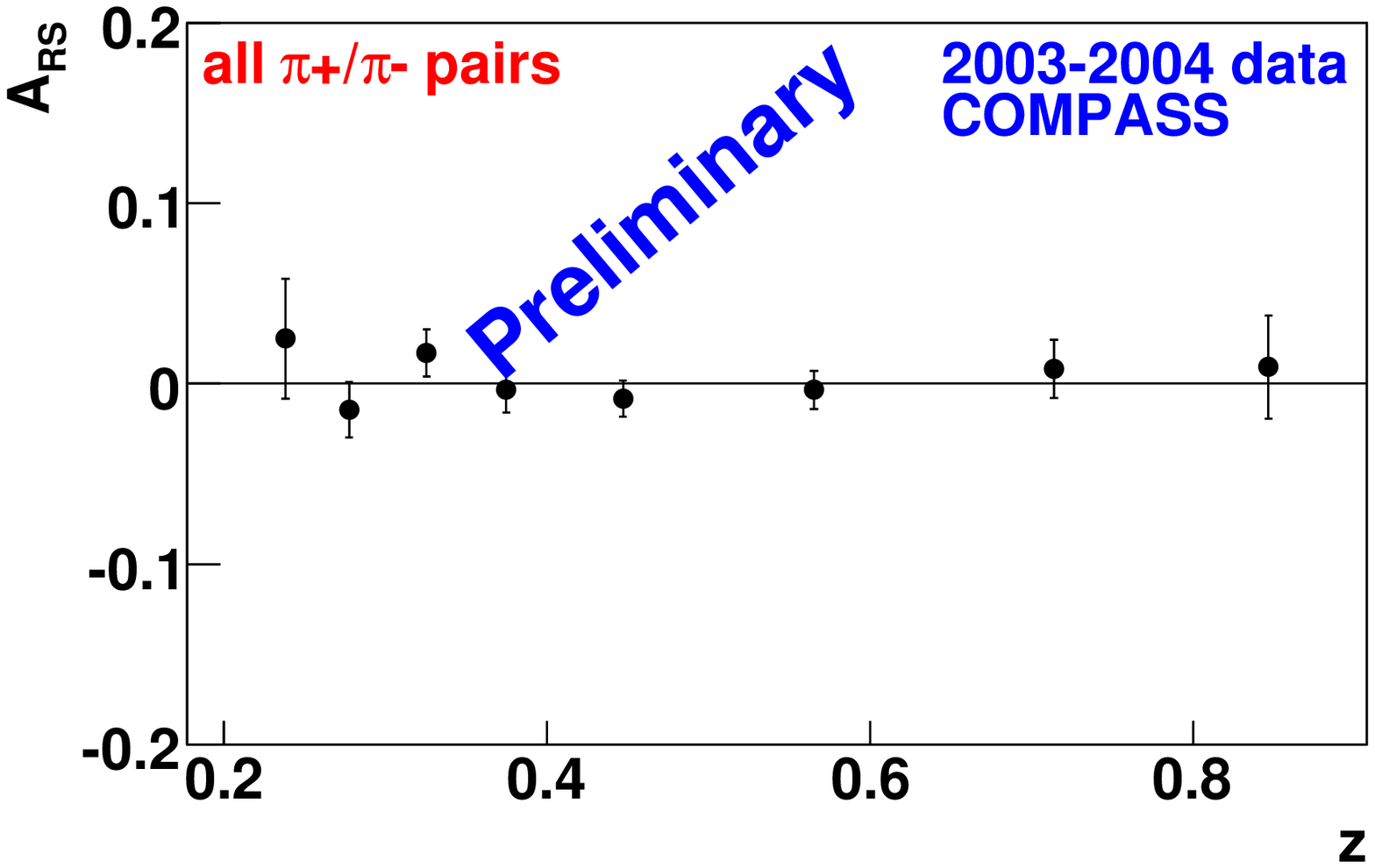}\hspace{-0.5cm}
\includegraphics[width=0.36\textwidth, height=3.1cm]{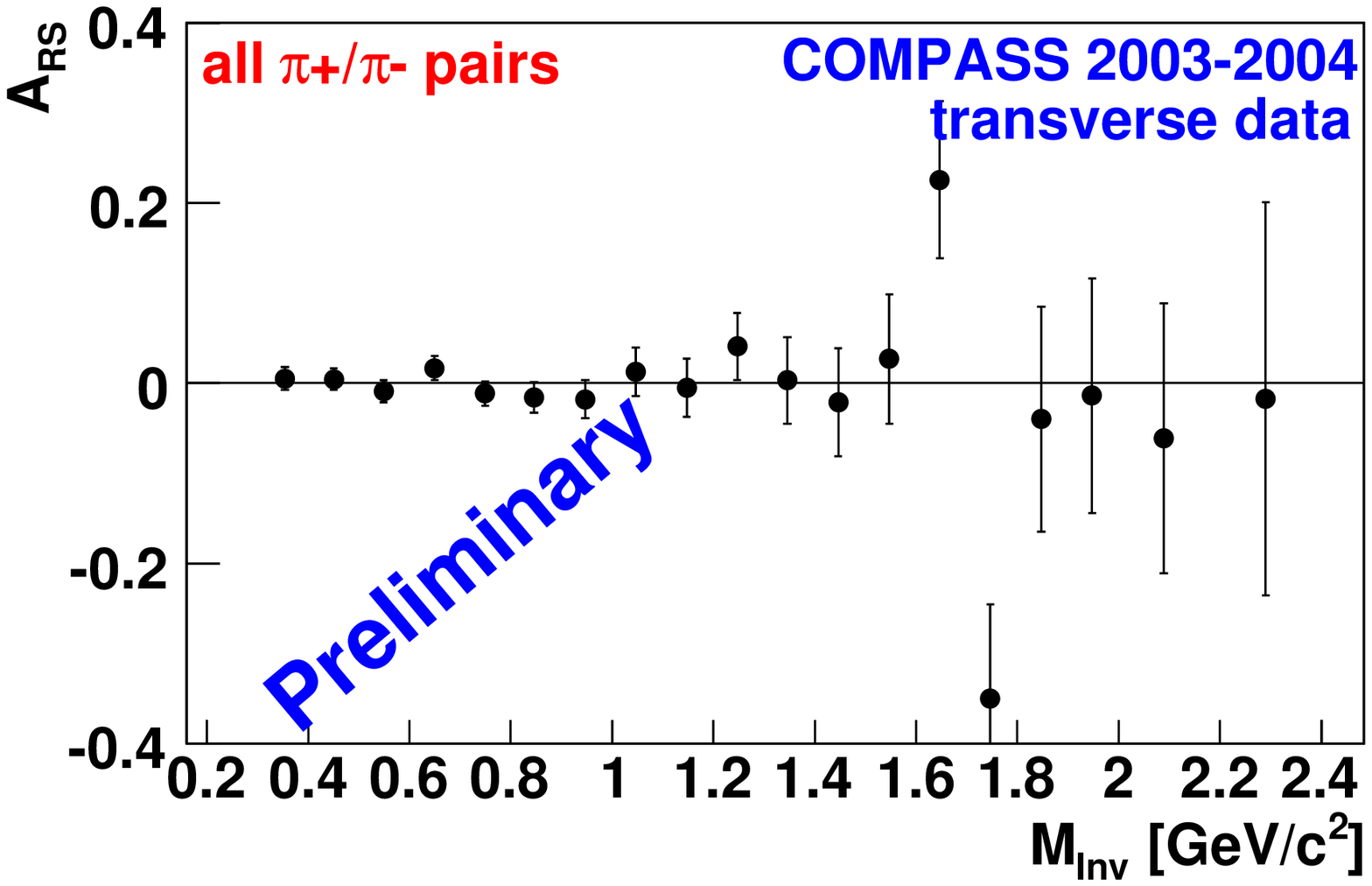}\hspace{-0.5cm}}
\vspace{-0.1cm}\centerline{
\includegraphics[width=0.36\textwidth, height=3.1cm]{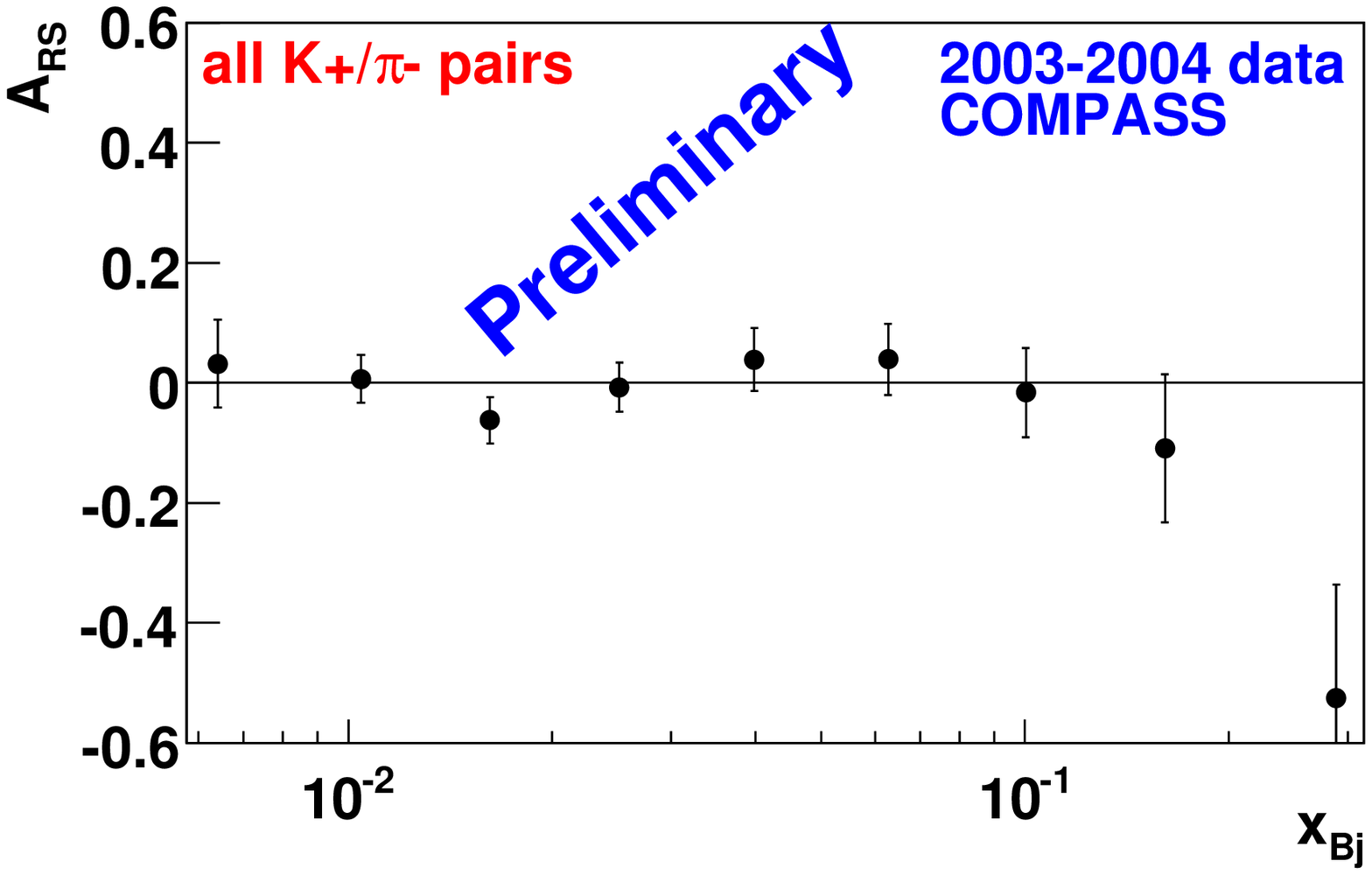}\hspace{-0.5cm}
\includegraphics[width=0.36\textwidth, height=3.1cm]{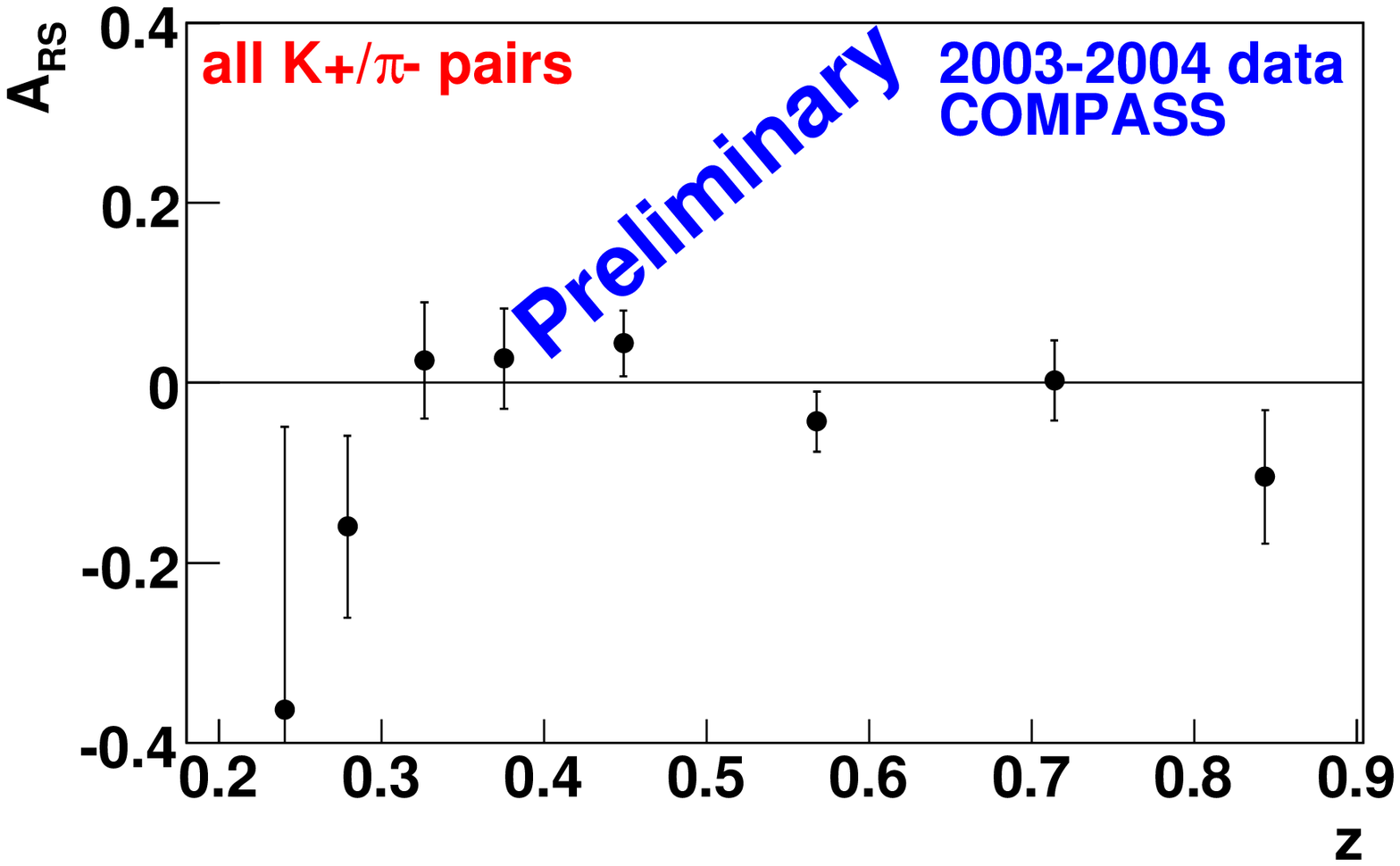}\hspace{-0.5cm}
\includegraphics[width=0.36\textwidth, height=3.1cm]{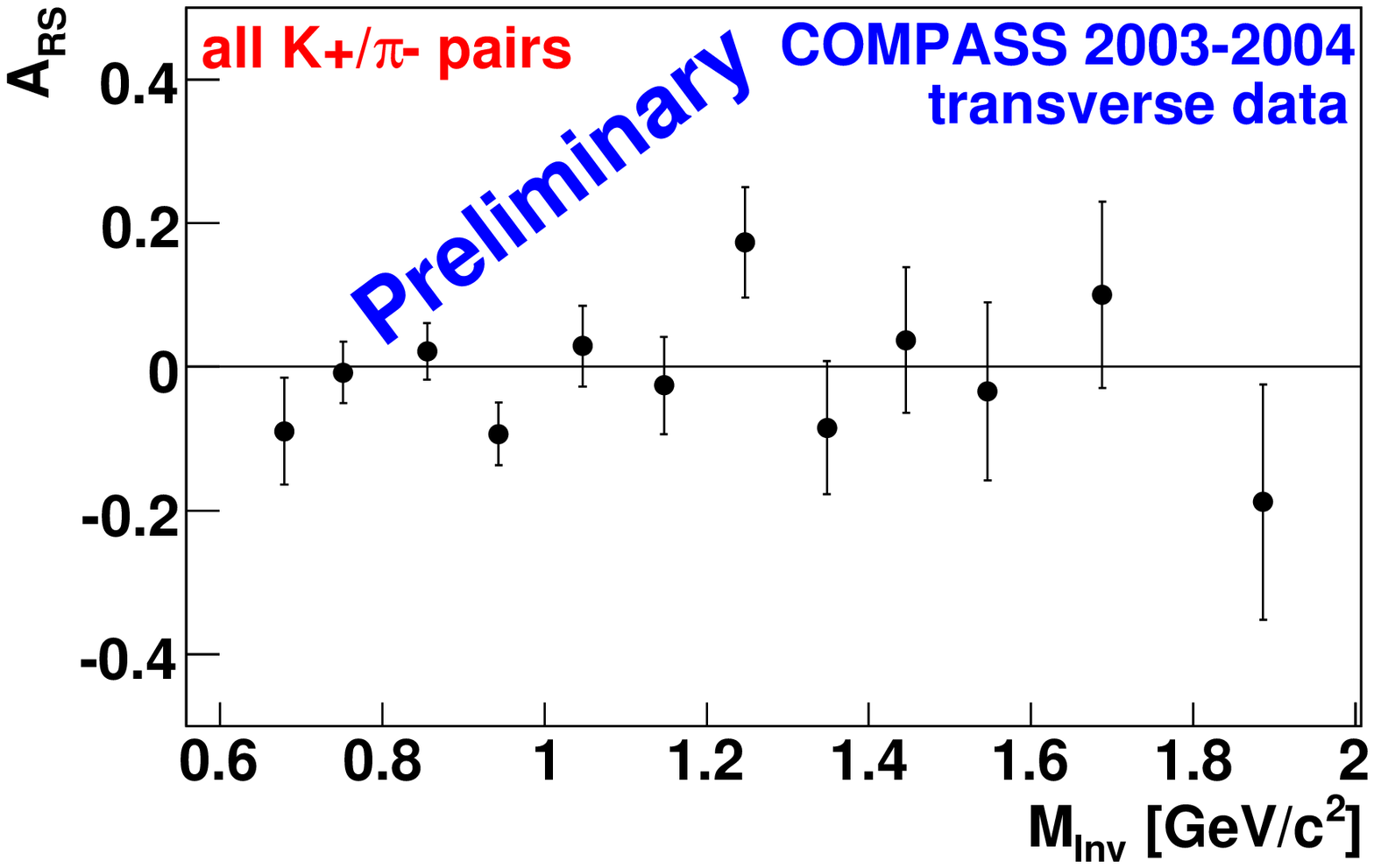}\hspace{-0.5cm}}
\vspace{-0.1cm}\centerline{
\includegraphics[width=0.36\textwidth, height=3.1cm]{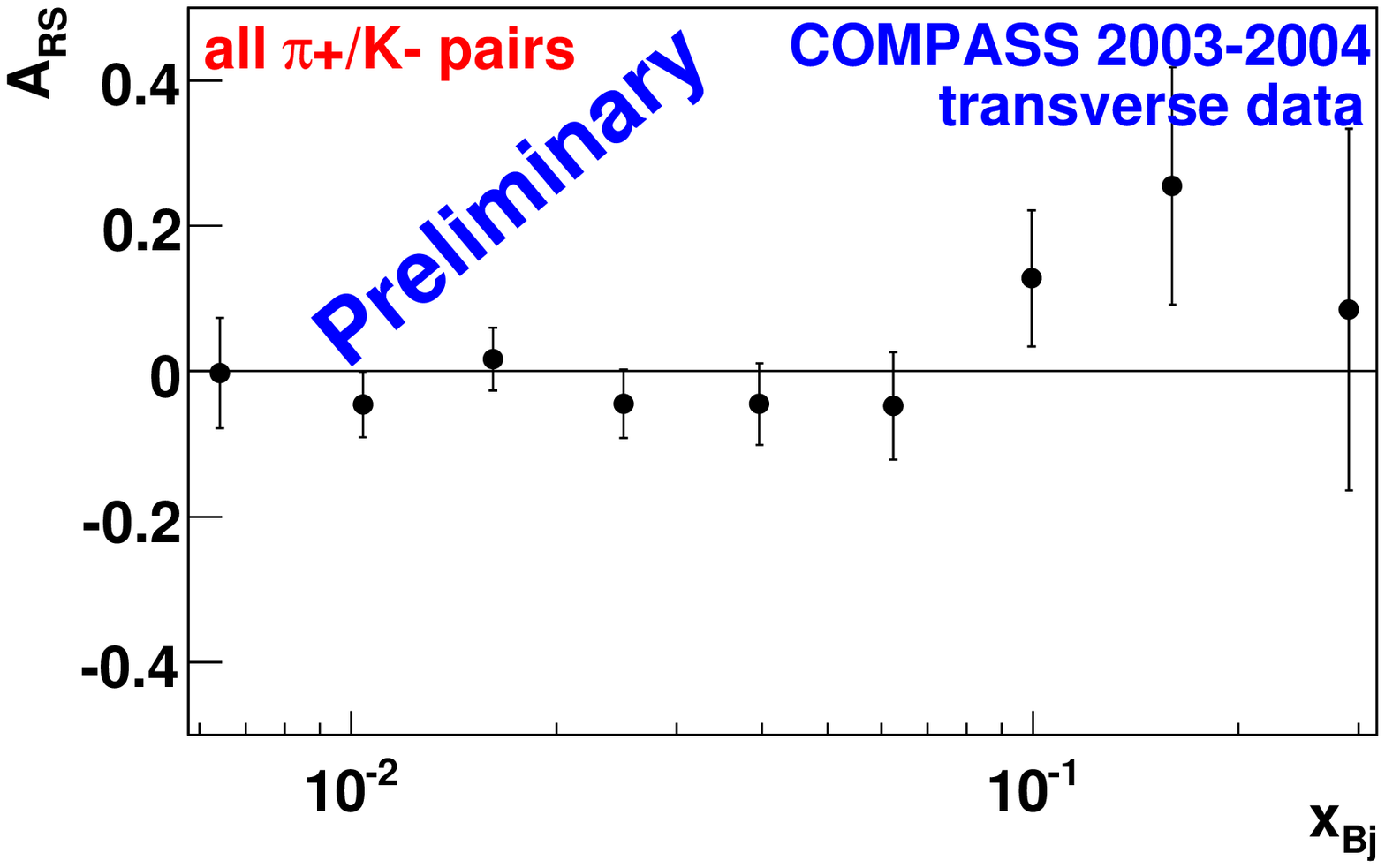}\hspace{-0.5cm}
\includegraphics[width=0.36\textwidth, height=3.1cm]{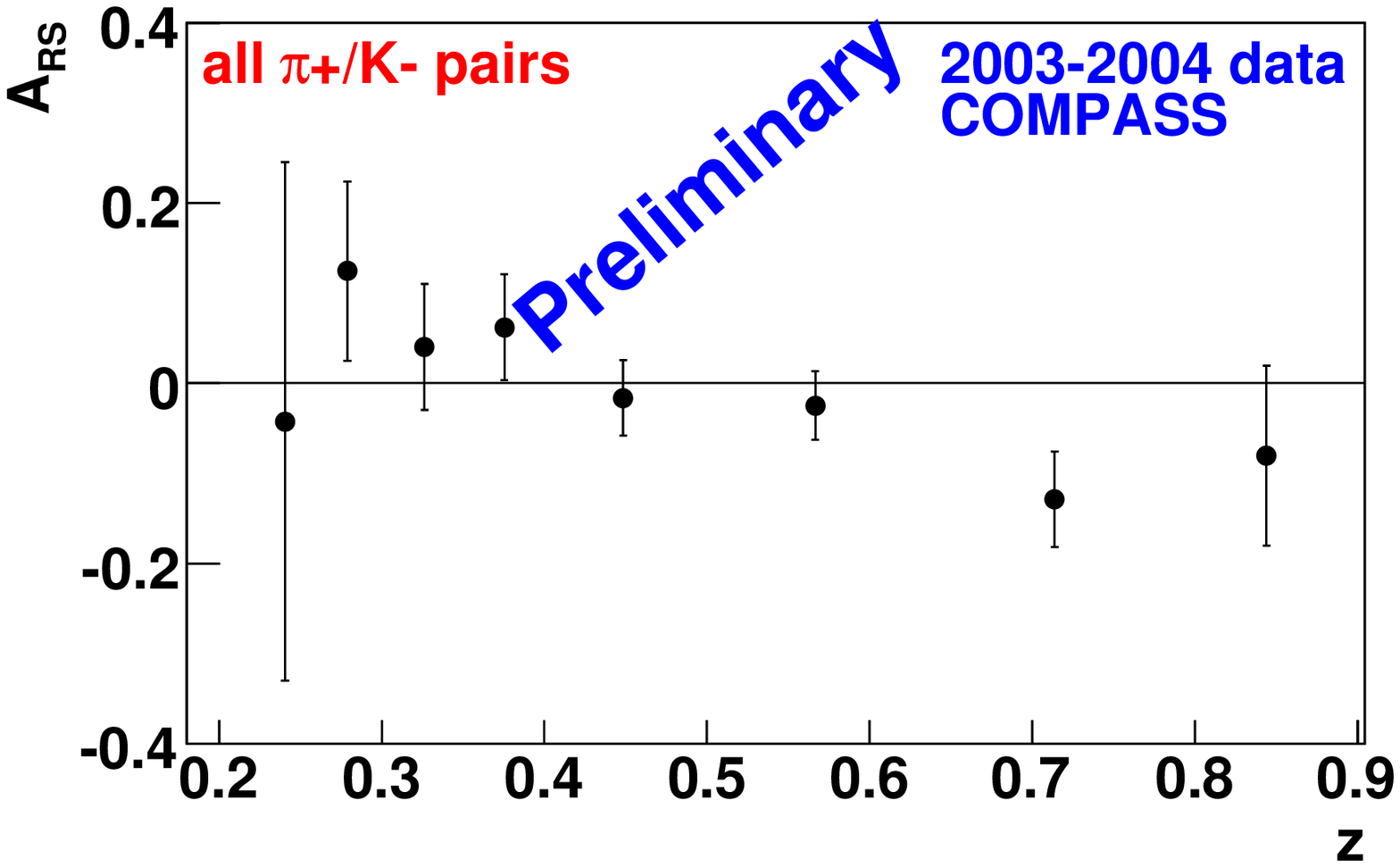}\hspace{-0.5cm}
\includegraphics[width=0.36\textwidth, height=3.1cm]{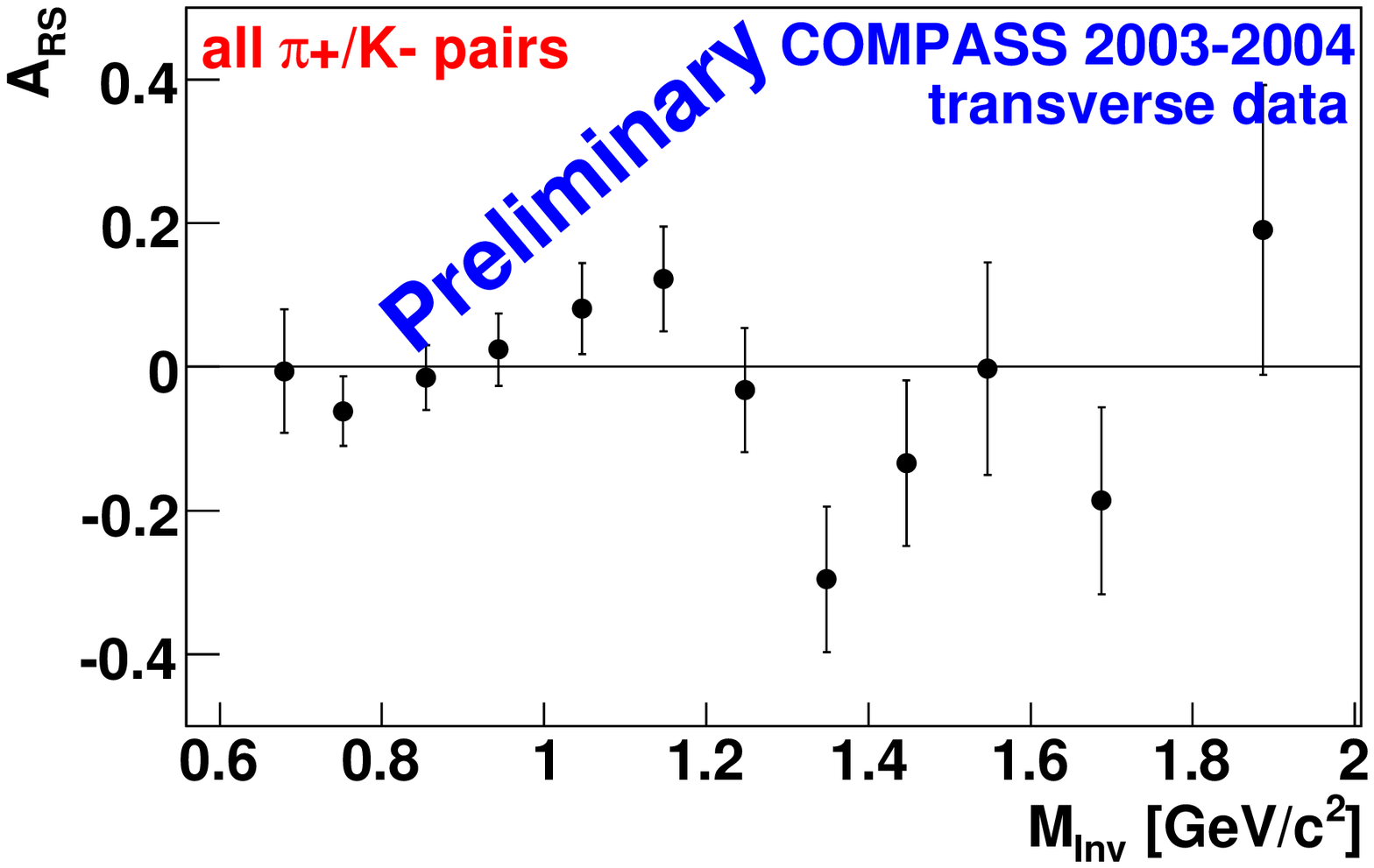}\hspace{-0.5cm}}
\vspace{-0.1cm}\centerline{
\includegraphics[width=0.36\textwidth, height=3.1cm]{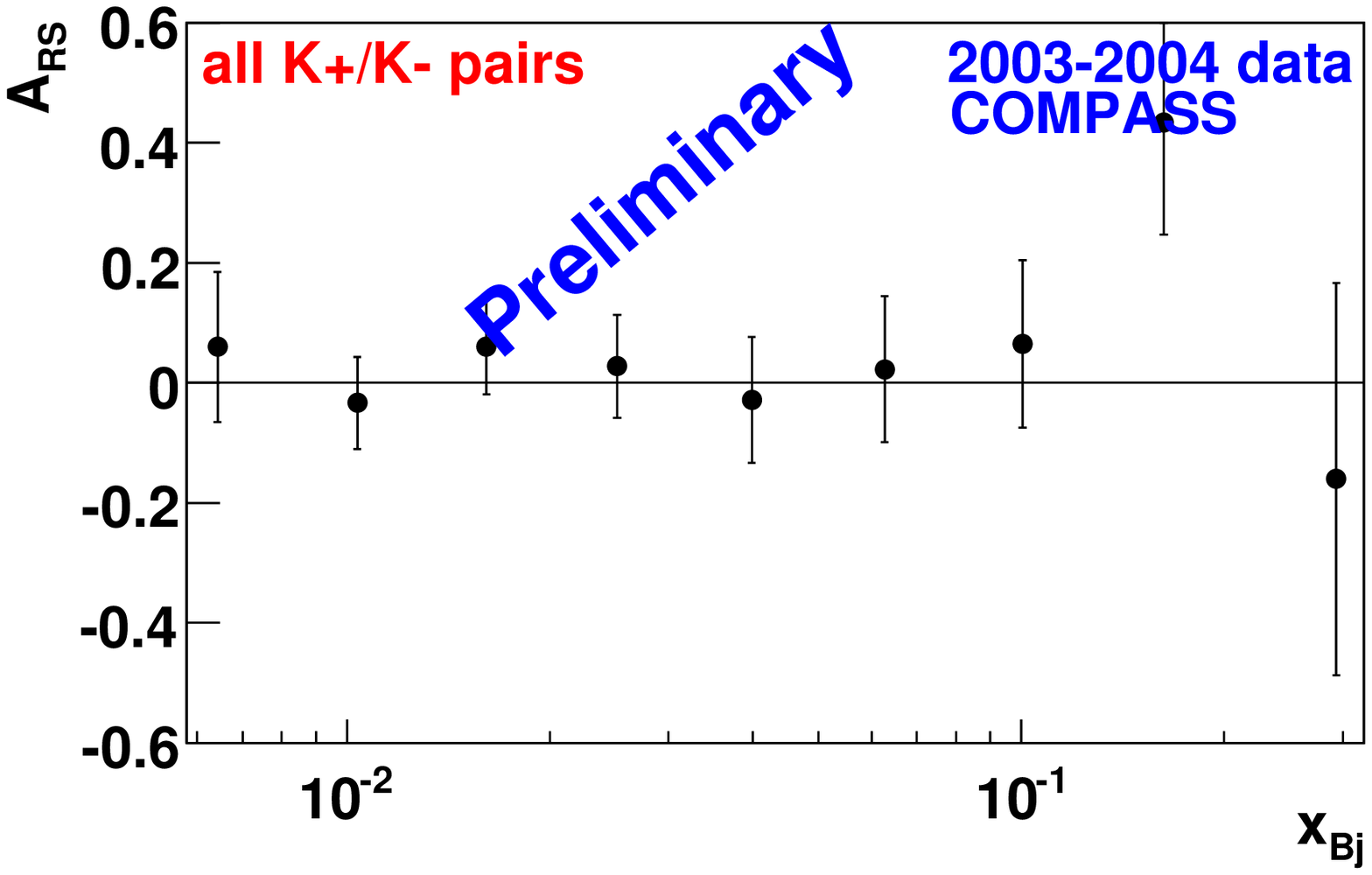}\hspace{-0.5cm}
\includegraphics[width=0.36\textwidth, height=3.1cm]{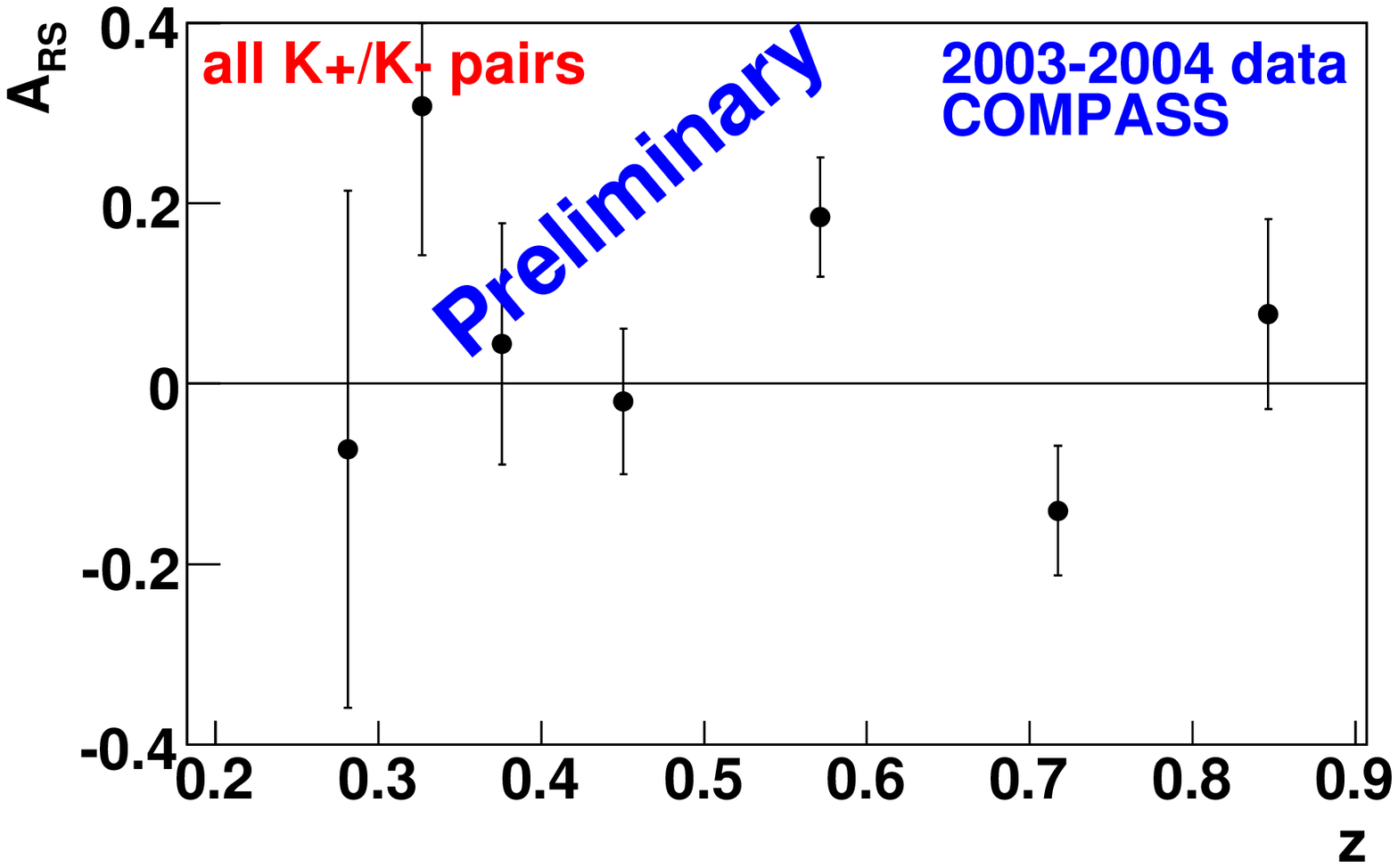}\hspace{-0.5cm}
\includegraphics[width=0.36\textwidth, height=3.1cm]{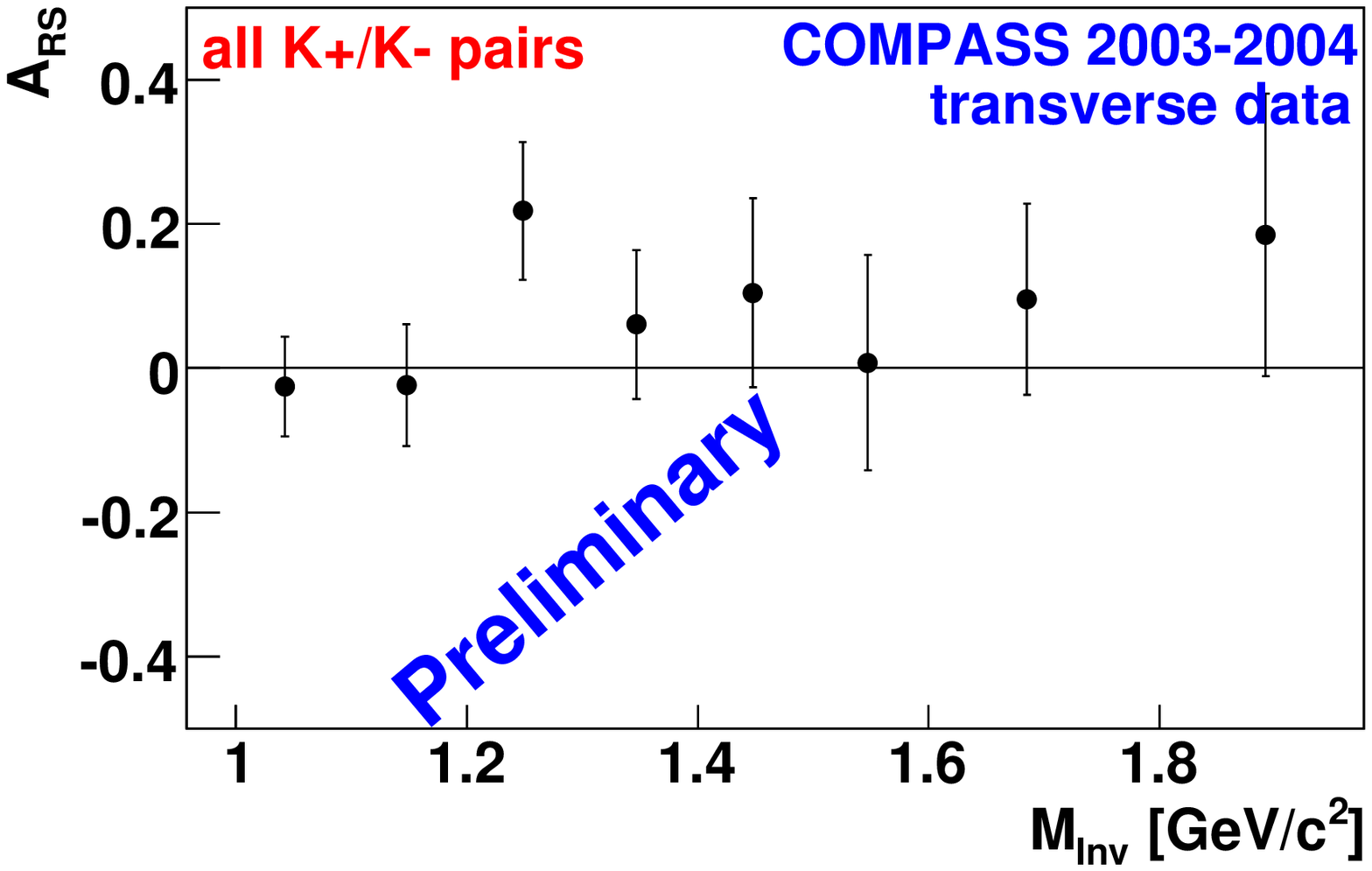}\hspace{-0.5cm}}
\vspace{-0.3cm}
\caption{Asymmetries $A_{RS}$ for identified hadron pairs: $\pi^+\pi^-$ (top
row), $K^+\pi^-$ (second row), $K^- \pi^+$ (third row) and $K^+K ^-$ (last row).
 The asymmetries are shown as a function of $x$ (left colum), $z$ (middle
 column) and $M_h$ (right column).}\label{Results}
 \vspace{-0.3cm}
\end{figure}

Figure \ref{Results} shows the preliminary results for the
target single spin asymmetry $A_{RS}(x, y, M_h)$ for identified hadron
pairs. In the first row, the asymmetry for $\pi^+\pi^-$ is displayed, in the
second for $K^+\pi^-$ pairs, in the third for $\pi^+K^-$ and in the last row
for $K^+K^-$ pairs. The asymmetries are plotted  as a function of $x$, $z$ and
$M_h$. The measured asymmetries  are very small and compatible with zero
within the statistical precision of the data points.  They do not show a
significant dependence on the kinematic variables $x$ and $z$ and on the hadron
pair invariant mass  $M_h$.

\section{Discussion}  

In several theoretical models, predictions have been made for the measured
asymmetries $A_{RS}(x, z, M_h)$ for pions or unidentified hadrons on a
deuteron target \cite{Bianconi, Bacchetta}. The expected values of the
asymmetry are generally small and below $1\%$. The small signal is attributed
in these calculations to a partial cancellation of the asymmetries originating
from scattering on $up$ and $down$ quarks of the proton and neutron in the
isoscalar deuteron target. In 2007, COMPASS is taking data with a transversely
polarized proton target, where the asymmetries are expected to be larger
\cite{Bacchetta}.
Together with the deuteron data presented here, a separation of the asymmetries
originating from $up$ and $down$ quarks shall then be possible.

\begin{footnotesize}

\end{footnotesize}



\begin{thebibliography}{99}
\bibitem{url} Slides:  
\verb$http://indico.cern.ch/contributionDisplay.py?contribId=165&sessionId=4&confId=9499$
\bibitem{Bressan} A. Bressan (COMPASS) and A. Kotzinian (COMPASS), these proceedings.
\bibitem{COMPASS} V.Yu. Alexakhin {\it et~al.} [COMPASS collaboration] Phys.
Rev. Lett. {\bf 94}, 202002 (2005) \newline and E.S. Ageev {\it et~al.} [COMPASS collaboration] Nucl. Phys. 
{\bf B765}, 31 (2007).
\bibitem{Collins} J.R. Collins, S.F. Heppelmann and G.A. Ladinsky, Nucl. Phys. 
{\bf B420}, 565 (1994).
\bibitem{Artru} X. Artru and J.C. Collins, Z. Phys.  {\bf C69}, 277 (1996).
\bibitem{Jaffe} R.L. Jaffe, X. Ji and J. Tang, Phys. Rev. Lett. {\bf 80}, 1166
(1998).
\bibitem{Bianconi} A. Bianconi, S. Boffi, R. Jakob and M. Radici, Phys. Rev. 
{\bf D62}, 034008 (2000).
\bibitem{Radici} M. Radici, R. Jakob and A. Bianconi, Phys. Rev. Lett {\bf D65},
074031 (2002).
\bibitem{Bacchetta} A. Bacchetta and M. Radici, Phys. Rev. {\bf D67}, 094002
(2003), Phys. Rev. {\bf D69}, 074026 (2004) and Phys. Rev. {\bf D74} 114007 (2006).
\bibitem{Artru2} X. Artru, hep-ph/0207309 (2002).
\bibitem{Experiment} P. Abbon {\it et~al.} [COMPASS collaboration]
hep-ex/0703049, accepted by NIM {\bf A}.



\end{thebibliography}
\end{document}